\pdfoutput=1 

\documentclass[11pt]{article}

\usepackage[letterpaper,margin=1.1in]{geometry}

\usepackage[T1]{fontenc}
\usepackage[utf8]{inputenc}     

\usepackage{amsmath}
\usepackage{amssymb}

\usepackage{newtxtext}
\usepackage{newtxmath}

\usepackage{graphicx}
\usepackage{empheq}             
\usepackage{caption}
\usepackage{float}
\usepackage[colorlinks=true,allcolors=blue,breaklinks=true]{hyperref}
\usepackage{url}
\providecommand{\JournalTitle}[1]{#1}

\providecommand{\bmsection}[1]{\medskip\par\noindent\textbf{#1.}\ \ignorespaces}

\begin{document}

\title{Angular resolution limit of augmented reality waveguides from guided mode theory}

\author{Guohua Wei\thanks{\texttt{gwei@meta.com, gwei2026@gmail.com}}\\
        \small Reality Labs, Meta, Redmond, Washington 98052, USA}

\date{}                         
\maketitle

\begin{abstract}
We derive the angular resolution limit of an augmented-reality (AR) waveguide combiner from guided mode theory. A slab of finite thickness supports only a discrete set of total-internal-reflection (TIR) angles, giving an angular resolution that worsens with thinner waveguides and smaller TIR angles. At 500~$\mu$m thickness the resolution is worse than the 1 arcmin (60 ppd) acuity target over most of the usable TIR range, and reaching 1 arcmin for TIR < 42$^\circ$ requires a substrate thicker than 1 mm. The limit is washed out by a source linewidth above $\sim$0.1~nm or a thickness variation above $\sim$100~nm across the pupil-replication period, but only partially by pupil truncation and not at all by out-coupling loss. LED- and micro-LED-based displays are therefore unaffected, owing to their broad spectra, whereas a narrow-linewidth laser-based display in a thin, uniform waveguide is subject to this limit.
\end{abstract}

\section{Introduction}
\label{sec:intro}

In augmented reality (AR) near-eye displays, the waveguide combiner that replicates the
exit pupil of a light engine across the eyebox is typically a transparent slab with a
refractive index chosen to support a target field of view (FOV): a higher index covers a
larger FOV. Glasses with visible-range refractive indices about $2.0$ are now widely
available for AR applications~\cite{schott,corning}, and materials of extraordinarily high
index, such as SiC, are being exploited for AR eyewear as well, such as the Meta Orion AR glasses ~\cite{orion}.

AR waveguides are manufactured over a range of thicknesses, typically a few hundred micrometres. Thickness impacts the pupil-replication density, the in-coupling and out-coupling efficiency, and the eyebox uniformity~\cite{levola,review,kress2020,coupler}, while thinner substrates are preferred for a lighter, more compact form factor and for higher replication density. In this article we show that thickness additionally imposes a limit on \emph{angular} resolution: a slab of finite thickness supports only a discrete set of TIR angles, so the propagation directions available to the combiner are quantized rather than continuous.

This effect is a guided mode resonance phenomenon. It is therefore closely related to the coherent pupil-replication interference already known to produce brightness non-uniformity and banding in laser-illuminated waveguide displays~\cite{shi2025}. Here, we derive the angular resolution from the guided mode theory of a slab waveguide, and establish the source-linewidth and thickness-uniformity criteria that determine whether the limit is observable at all. We further show that pupil truncation partially relaxes the requirement, and removes the limit only when the pupil replication period approaches the eye pupil diameter, so that a single replica reaches the eye. We also show that the grating out-coupling loss does not remove it at any realistic out-coupling efficiency.

Mode theory describes systems in which electromagnetic waves are confined in one or more dimensions. An AR waveguide can be modelled as a slab waveguide confined in the out-of-plane direction. We first recover the TE guided mode conditions for a symmetric slab (Sec.~\ref{sec:modes}), then convert the mode spacing into an angular resolution $\delta\theta$ (Sec.~\ref{sec:resolution}), and finally derive the source-linewidth and thickness-uniformity criteria under which the quantization is washed out (Sec.~\ref{sec:washout}). A geometric-optics round-trip phase argument reproduces the same mode spacing (Sec.~\ref{sec:geometric}).

Throughout we adopt the following assumptions and discuss their validity in Sec.~\ref{sec:discussion}: the slab is infinite and translationally invariant in the propagation plane; the in-coupling is approximately a plane wave of a single linear polarization (TE); and the illumination is coherent both spatially and temporally over the round-trip optical path. We also show how the limit can be relaxed by moving away from any of these assumptions.

\section{Guided mode theory}
\label{sec:modes}

\begin{figure}
  \begin{center}
    \includegraphics[width=0.49\linewidth]{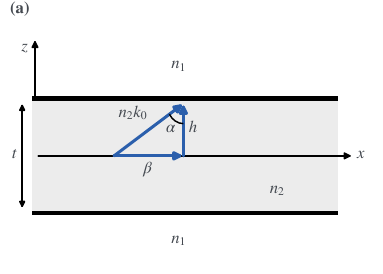}
    \includegraphics[width=0.49\linewidth]{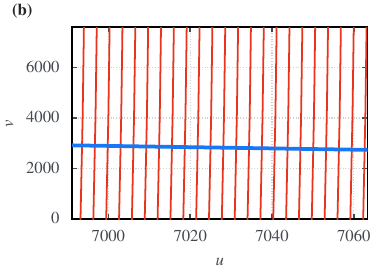}
  \end{center}
  \caption{(a) A slab waveguide with thickness of $t$. The TIR angle $\alpha$ is measured from
  the surface normal $\hat{z}$, so that $h = n_2 k_0 \cos\alpha$ and
  $\beta = n_2 k_0 \sin\alpha$. (b) Symmetric TE modes for TIR angle from $30^{\circ}$ to $31^{\circ}$. $n_1 = 1$, $n_2 = 2.7$, $t = 500~\mu$m, $\lambda_0 = 520$~nm.}
  \label{fig:1}
\end{figure}

Consider a waveguide as shown in Fig.~\ref{fig:1} with a thickness of $t$ and an index
$n_2$, air cladding on both sides. A mode must have a unique propagation constant (usually
noted as $\beta$) and well-defined amplitude at each point of space and time\cite{yariv}. The wave equation follows:
\begin{align}
\left(\frac{\partial^{2}}{\partial y^{2}}+\frac{\partial^{2}}{\partial z^{2}}\right) \vec{E}(y ,z)+\left(k_{0}^{2} n^{2}-\beta^{2}\right) \vec{E}(y,z)=0
\label{eq:wave}
\end{align}
$\beta$ is the propagation constant, or effective k-vector of the mode. Since the slab is
also invariant along $y$, $\partial^{2}/\partial y^{2}=0$ and the field depends on $z$
only.

Assuming TE modes in the waveguide, so that the electric field only has a $y$ component:
\begin{equation}
\vec{E}=E_{y}(x, z, t) \hat{y}=E(z) \exp [i(\omega t-\beta x)]\,\hat{y}
\end{equation}

Define
\begin{equation}
\begin{cases}
h^{2}=n_{2}^{2} k_{0}^{2}-\beta^{2} \\
q^{2}=\beta^{2}-n_{1}^{2} k_{0}^{2}
\end{cases}
\end{equation}
the boundary conditions at the interfaces of $z=\frac{1}{2} t$ and $z=-\frac{1}{2} t$
yield the solutions of the modes. For symmetric (even) modes
\begin{equation} \label{eqn12}
    h \tan \left(\frac{1}{2} h t\right)=q.
\end{equation}
For antisymmetric (odd) modes
\begin{equation} \label{eqn13}
    h \cot \left(\frac{1}{2} h t\right)=-q.
\end{equation}

Let $u=\frac{1}{2} h t, \,v=\frac{1}{2} q t$,
\begin{equation}
\begin{aligned}\label{eqn15}
u^{2}+v^{2} &=\frac{1}{4} t^{2}\left(h^{2}+q^{2}\right)=\frac{1}{4} t^{2}\left(n_{2}^{2}-n_{1}^{2}\right) k_{0}^{2}=\left(\frac{\pi t}{\lambda_{0}}\right)^{2}\left(n_{2}^{2}-n_{1}^{2}\right) \\
&=\mathbf{V}^{2},
\end{aligned}
\end{equation}

Equation (\ref{eqn12}) becomes
\begin{equation}\label{eqn16}
    u \tan (u)=v .
\end{equation}

Equation (\ref{eqn13}) becomes
\begin{equation}\label{eqn17}
    v=-u \cot (u).
\end{equation}

Solving equations of Eq.~(\ref{eqn15}) and Eq.~(\ref{eqn16}) gives the mode conditions for symmetric TE modes. Likewise, equations of Eq.~(\ref{eqn15}) and Eq.~(\ref{eqn17}) give the mode solutions for antisymmetric TE modes.
The value of $\mathbf{V}$ as a function of $t/\lambda_0$ determines the number of guided
modes for a given wavelength. Since $u=\frac{1}{2}ht$, and $h$ is effectively the $z$ component of the guided modes as $h=\sqrt{n_2^2k_0^2-\beta^2}=n_2k_0\cos(\alpha)$ where $\alpha$ is the TIR angle, this
means that for each of these TE modes there is a different TIR angle $\alpha$ from the
geometric optics perspective.

When $t\gg\lambda_0$, such as for $t$ = 500~$\mu$m and $\lambda_0$ = 520~nm, the number of
guided modes increases to thousands for all the supported TIR angles. For a specific TIR
angle range as shown in Fig.~\ref{fig:1}(b), for $\alpha$ from $30^\circ$ to
$31^\circ$, we can see over 20 symmetric modes (a similar number of antisymmetric modes
are not shown). The finite number of modes limits the number of FOV angles that can be
supported for this wavelength.

\section{Waveguide optical angular resolution}
\label{sec:resolution}

For AR waveguides, considering a simple case where the input coupling angle is $\theta$,
an input grating of period $\Lambda$ (grating vector magnitude $G = 2\pi/\Lambda$, so that
$G/k_0 = \lambda_0/\Lambda$), and considering only one first-order in-coupling, we get

\begin{equation}\label{gratingeq}
n_{1} \sin \theta +\frac{\lambda_{0}}{\Lambda} =n_{2} \sin \alpha
\end{equation}
which gives

\begin{equation}
\delta \theta=\frac{n_{2}}{n_{1}} \frac{\cos \alpha}{\cos \theta} \cdot \delta\alpha
\label{eq_delta-theta}
\end{equation}

For guided modes, since
\begin{equation}\label{eq_cosa}
\cos \alpha=\frac{h}{n_{2} k_{0}} =\frac{2 u}{t n_{2} k_{0}}
\end{equation}
 which means,
\begin{equation}
\delta \alpha =-\frac{2}{t n_{2} k_{0}\sin \alpha } \delta u
\end{equation}
substitute it for Eq.~(\ref{eq_delta-theta}) we get,

\begin{equation}
\begin{aligned}
\delta \theta &=\frac{n_{2}}{n_{1}} \cdot \frac{\cos \alpha}{\cos \theta} \cdot \delta \alpha \\
&=-\frac{n_{2}}{n_{1}} \cdot \frac{1}{\cos \theta} \cdot \frac{\cos \alpha}{\sin \alpha} \cdot \frac{2}{t n_{2} k_{0}} \delta u \\
\end{aligned}
\end{equation}

When $t$ is large, the mode spacing for TIR angles can be approximated as
$\delta u = \pi/2$ (considering both symmetric and antisymmetric modes,
$v=u\tan(u)$ and $v=-u\cot(u)$, and only positive values are valid), so that

\begin{equation}\label{eq-f}
 \delta \theta \approx\frac{1}{n_{1}\cos \theta} \cdot \frac{\lambda_{0}}{2t \tan \alpha}
\end{equation}
with $n_{1} = 1$ and normal in-coupling angles $\theta = 0$, the angular resolution reduces to
\begin{equation}\label{eq:theta}
 \delta \theta \approx \frac{\lambda_{0}}{2t \tan \alpha}
\end{equation}

Equation~(\ref{eq:theta}) is the angular resolution of an ideal AR waveguide. It carries no explicit dependence on the substrate index: at a given TIR angle $\delta\theta$ is fixed by thickness and wavelength alone, and $n_2$ enters only through the TIR angle that the in-coupler produces. It shows that thinner waveguides sample the FOV more coarsely, and that the limit is worse at \emph{small} TIR angles because $\delta\theta \propto 1/\tan\alpha$. The worst case in any design is therefore the ray closest to the critical angle $\alpha_c = \arcsin(n_1/n_2)$.

\begin{figure}
  \begin{center}
     \includegraphics[width=0.49\textwidth]{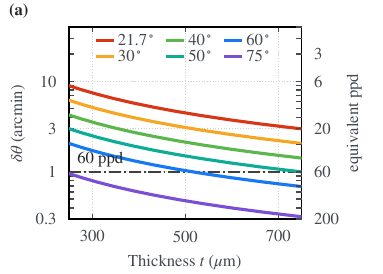}\hfill
     \includegraphics[width=0.49\textwidth]{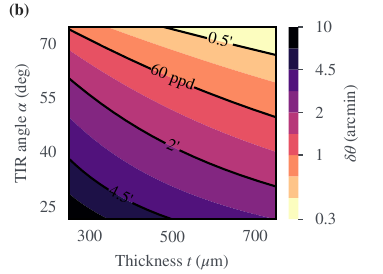}
    \caption{Angular resolution $\delta\theta$ versus waveguide thickness and TIR angles for TIR from the critical angle $\alpha_c = 21.7^\circ$ to $75^\circ$ with $n_1 = 1$, $n_2 = 2.7$, $\lambda_0 = 520$~nm. (a) shows a few representative TIR angles and (b) shows 2D mapping of the entire TIR angle range.}
    \label{fig:angularresolution}
  \end{center}
\end{figure}

Fig.~\ref{fig:angularresolution} plots Eq.~(\ref{eq:theta}) against thickness for TIR from the critical angle to $75^\circ$ with $\lambda_0 = 520$~nm. Normal human visual acuity ($20/20$) corresponds to a minimum angle of resolution of about $1$~arcmin (60~ppd), and a recent study~\cite{acuity} showed that humans can resolve even higher ppd under certain conditions. As shown by the plots, the guided mode theory limit indicates that even with a $500~\mu$m thick substrate, most of the TIR angles have a mode-spacing-limited resolution above 1 arcmin. For example, at the TIR angle actually produced by a $\Lambda = 299.6$~nm in-coupler at normal incidence, $\alpha = 40^\circ$, the $500~\mu$m waveguide already gives $2.13$~arcmin. Increasing the thickness to $750~\mu$m still leaves the limit above 1 arcmin for most of the TIR angles (<50$^\circ$). The waveguide thickness required to reach $1'$ is listed in Table~\ref{tab:thresholds}. For TIR angles below about 42$^\circ$, the thickness is above 1~mm, which is unfavorable for a compact, lightweight AR display.

\begin{table}[htbp]
  \begin{center}
  \caption{Waveguide thickness required to reach $1$~arcmin (60~ppd) resolution, as a function of TIR angle. $n_1 = 1$, $\lambda_0 = 520$~nm, $\theta = 0$.}
  \label{tab:thresholds}
 \begin{tabular}{lr}
 \hline TIR angle $\alpha$ & $t$ required \\
 \hline $21.74^\circ$ (critical for $n_2 = 2.7$) & \textbf{2242\,$\mu$m} \\
  $40^\circ$ & \textbf{1065\,$\mu$m} \\
  $50^\circ$ & 750\,$\mu$m \\
  $60^\circ$ & 516\,$\mu$m \\
  $75^\circ$ & 239\,$\mu$m \\
  \hline
  \end{tabular}
  \end{center}
\end{table}

The guided mode condition means that for a given wavelength $\lambda_0$, there is a set of TIR angles supporting guided modes in this slab waveguide, and this set of TIR angles is not continuous. In other words, for some in-coupling FOV angles $\theta$ (and equivalent out-coupling angles to the eyebox), the corresponding TIR angles are not guided modes so that they will not propagate in the waveguide. The finite angular space sampling sets a limit for the optical resolution.

\section{\texorpdfstring{Washout mechanisms: source linewidth and thickness \mbox{uniformity}}{Washout mechanisms: source linewidth and thickness uniformity}}
\label{sec:washout}

One way to relieve the limitation described by Eq.~(\ref{eq:theta}) is to have a meaningful
linewidth for the light sources utilized by the display engine. This is a particular
concern for laser-based displays, where the linewidth can be very narrow\cite{shi2025}.

For any given TIR angle, observing from the numerical solutions above, one can see that
when $\Delta u>\pi/2$ over a given range $\Delta \lambda$, there is always at least one
guided mode for that TIR angle, which removes the discrete TIR angles and the angular resolution limit.

From Eq.~(\ref{eq_cosa}), we know
\begin{equation}\label{eq-u-alpla}
 u=\frac{1}{2}tn_2k_0\cos\alpha
\end{equation}
so that
\begin{equation}\label{eq-deltau}
 \delta u=-\left(\frac{n_2t\pi\cos\alpha}{\lambda_0^2}+\frac{\tan\alpha}{\Lambda n_2}\frac{n_2t\pi}{\lambda_0}\right)\delta\lambda
\end{equation}
by exploiting the grating Eq.~(\ref{gratingeq}). The second term in
Eq.~(\ref{eq-deltau}) originates from the grating dispersion and is not negligible. With $|\delta u| = \pi/2$, we get

\begin{equation}\label{eq:dlambda}
 \delta \lambda = \frac{1}{2t}\frac{\lambda_0^2\Lambda}{n_2\Lambda\cos\alpha+\lambda_0\tan\alpha}
\end{equation}

\begin{figure}
  \begin{center}
    \includegraphics[width=0.49\textwidth]{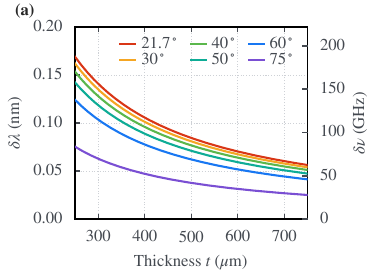}\hfill
    \includegraphics[width=0.49\textwidth]{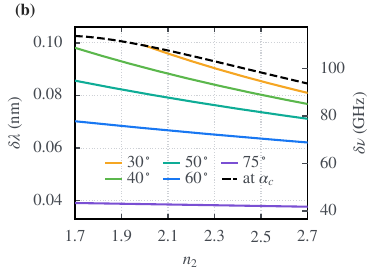}
    \caption{Source linewidth required for continuous TIR-angle coverage, versus (a) waveguide thickness and (b) waveguide refractive index at thickness $t = 500~\mu$m. The right axis gives the equivalent frequency width at $\lambda_0 = 520$~nm, the dashed curve gives $\delta \lambda$ at the critical angles, using $\Lambda =$ 300~nm, $\theta = 0$.}
    \label{fig:linewidth}
  \end{center}
\end{figure}

This is the linewidth requirement for the display light source to ensure continuous TIR
angle coverage so that any FOV angle can be supported by the waveguide guided modes, and
it is plotted against thickness in Fig.~\ref{fig:linewidth}(a). The grating
dispersion term relaxes the requirement on linewidth broadening. For example, at $\lambda_0 = 520$~nm and $\alpha = 40^\circ$ the two terms are $7.65$ and $5.37$~nm$^{-1}$ respectively, so that the grating dispersion reduces the linewidth requirement by a factor of $1.70$ relative to the bare etalon
free spectral range $\lambda_0^2/(2 n_2 t \cos\alpha)$, which gives $0.131$~nm at
$t = 500~\mu$m against $0.077$~nm from Eq.~(\ref{eq:dlambda}). The grating dispersion also reduces the dependency on $n_2$ as shown in Fig.~\ref{fig:linewidth}(b): the linewidth requirement changes slowly with refractive index.

It is instructive to look at Eq.~(\ref{eq:dlambda}) in frequency and coherent length units. At $t = 500~\mu$m, $\delta\lambda = 0.077$~nm corresponds to
$\delta\nu = 85$~GHz and to a coherence length $\lambda_0^2/\delta\lambda = 3.5$~mm, the
latter being comparable to the round-trip optical path in the slab. The criterion is
therefore simply that the source coherence length be shorter than the round trip optical path. A
single-longitudinal-mode laser diode, with a linewidth well below $1$~GHz, is well within
this constrained regime; a multi-longitudinal-mode laser diode spanning more than $85$~GHz or $\sim 0.1$~nm linewidth, and
any LED or micro-LED based display engine with a linewidth above $10$~nm, is far outside of this constraint.

\begin{figure}
  \begin{center}
    \includegraphics[width=0.6\textwidth]{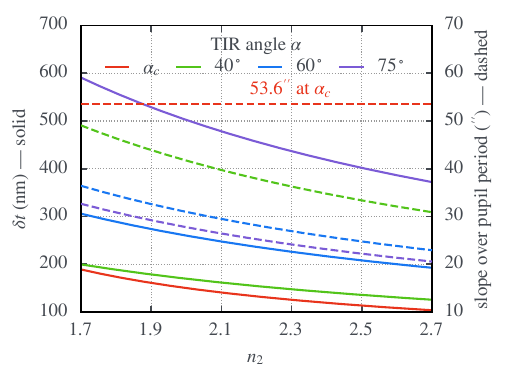}
    \caption{Local thickness variation requirement for mode washout. The right axis gives the equivalent slope across the replicated pupil period in the propagating direction for $\lambda_0 = 520$~nm, and $t = 500~\mu$m.}
    \label{fig:dt}
  \end{center}
\end{figure}

Waveguides typically possess thickness variation. What matters for the phenomenon we discuss here is the short range thickness variation, sometimes referred to as local slope or wedge. Differentiating $u = \pi t n_2 \cos\alpha / \lambda_0$ with
respect to $t$ and requiring $|\delta u| = \pi/2$ gives
\begin{equation}
\delta t = \frac{\lambda_0}{2 n_2 \cos\alpha}
\label{eq:dt}
\end{equation}
which is independent of $t$.
Fig.~\ref{fig:dt} plots Eq.~(\ref{eq:dt}) over the usable TIR range for $n_2$ from 1.7 to 2.7 and $\lambda_0 = 520$~nm. A thickness variation exceeding 100~nm is required over the full range of TIR angles and refractive indices considered. Using the previous example, for $\alpha = 40^\circ$, $\delta t = 126$~nm. With a $500~\mu$m thick waveguide, this indicates a wedge of  $30.9$~arcsec over a $0.84$~mm pupil replication footprint (pupil pitch). A thickness variation or wedge of this magnitude shifts the local mode comb by one full mode spacing, so the discrete TIR angles average out across the pupil and the resolution limit of Eq.~(\ref{eq:theta}) does not apply.

The criterion of Eq.~(\ref{eq:dt}) depends on the substrate index and on the TIR angle. Because $\delta t$ grows monotonically with $\alpha$, the lowest requirement for any substrate occurs at its critical angle $\alpha_c = \arcsin(n_1/n_2)$, where $\cos\alpha_c = \sqrt{1-n_1^2/n_2^2}$, and Eq.~(\ref{eq:dt}) reduces to
\begin{equation}
\delta t_{\min} = \frac{\lambda_0}{2\sqrt{n_2^2-n_1^2}} \label{eq:dtmin}
\end{equation}
a function of the substrate index alone. For $\lambda_0 = 520$~nm this is $189$~nm at $n_2 = 1.7$ and falls to $104$~nm at $n_2 = 2.7$: raising the index reduces the uniformity requirement both directly, through Eq.~(\ref{eq:dt}), and indirectly, by having shallower TIR angles at which $\delta t$ is smallest. High-index substrates are therefore favored on both counts.

The thickness uniformity here is not measured as an absolute thickness change but as a local wedge/slope, and the length over which that slope is taken is set by the pitch of successive pupil replicas $p = 2t\tan\alpha$, so the relevant quantity is $\delta t$ divided by $p$,

\begin{equation} \label{eq:slope}
\left|\frac{\mathrm{d}t}{\mathrm{d}x}\right| = \frac{\delta t}{2t\tan\alpha} = \frac{\lambda_0}{4 n_2 t \sin\alpha}
\end{equation}

Evaluated at the critical angle, where $\sin\alpha_c = n_1/n_2$, the index cancels identically and Eq.~(\ref{eq:slope}) becomes simply $\lambda_0/(4 n_1 t)$, equal to $53.6''$ for $\lambda_0 = 520$~nm, $n_1 = 1$ and $t = 500~\mu$m --- the same value for every substrate index and shown as the red dashed line in Fig.~\ref{fig:dt}. This invariance is a convenient result for specification purposes: the highest local non-uniformity (local wedge) requirement is fixed by wavelength and thickness, regardless of substrate material refractive index.

The left axis of Fig.~\ref{fig:dt} reads as an absolute thickness variation and a higher index is thus less demanding, from $189$~nm to $104$~nm at $\alpha_c$ for substrate index from 1.7 to 2.7. A similar trend holds for the local wedge versus index, as shown by the right axis: for instance, $49.1''$ at $n_2 = 1.7$ against $30.9''$ at $n_2 = 2.7$ for $\alpha = 40^\circ$. Also notice that for smaller TIR angles, the local thickness variation (slope) requirement is higher. This is because the pupil replication pitch contracts faster with smaller $\alpha$ than $\delta t$ does. A waveguide local uniformity specification is meaningful only when the aperture over which it is measured is explicitly defined.

It is worth noting that washing out the modes requires a local thickness variation exceeding $\delta t$, whereas image sharpness (MTF) requires the opposite. Every pupil replication deviates the pupil by twice the local slope, so the pupil replicas emerge as a fan of equally spaced exit angles when diffracted by the output grating and the wavefront error accumulates across the replicated pupils~\cite{wedge}, causing image blur. The effect is more severe under incoherent illumination where the replicas add in intensity and the fan superposes as laterally displaced images rather than interfering.

\section{Geometric explanation}
\label{sec:geometric}

This behavior can be understood from the geometric optics perspective with some
approximation. To have a guided mode in the waveguide, the phase accumulated over a
round-trip for a ray bouncing between the two interfaces must be an integer multiple of
$2\pi$, which means:

\begin{equation}
2 h t - 2\phi =m \cdot 2 \pi, \quad m\in \mathbf{Z} .
\end{equation}
where $\phi$ is the phase advance on each TIR, taken positive with a value in
$[0, \pi]$. Since $\phi$ varies slowly with $\alpha$ compared with $ht$ when
$t \gg \lambda$, the difference in $\phi$ between adjacent modes may be neglected and we
recover the mode spacing $\Delta u =\Delta(\frac{1}{2}ht) = \pi/2$, and hence
Eq.~(\ref{eq-f}).

\section{Discussion}
\label{sec:discussion}

The guided mode derivation assumes an ideal waveguide slab where the phase is stable at each bounce, and the mode is lossless. In addition, the mode quantization presumes an infinite, translationally invariant slab excited by a fully
coherent plane wave. A real waveguide has a few practical differences that can help relax this limitation.

First, the input pupil, eye pupil, and the grating footprints are finite. Successive pupil replicas emerge displaced laterally by one pupil period $p = 2t\tan\alpha$, and only those replicas that overlap within the eye pupil can interfere at the retina, so the number of interfering input pupil replicas is
\begin{equation}
N_{\rm rep} = \frac{D_{\rm eye}+D_{\rm ip}}{2t\tan\alpha}\,, \label{eq:Nrep}
\end{equation}
As an example, with eye pupil diameter $D_{\rm eye} = 3$~mm, and waveguide input pupil $D_{\rm ip} = 3$~mm, $t = 500~\mu$m and $\alpha = 40^\circ$, $N_{\rm rep}$ equals $7$ but with different weights. Assuming a uniform amplitude of $N_{\rm rep}$ replicas gives resonances of width $0.886/N_{\rm rep}$ of the free spectral range\cite{bornwolf}, hence
 \begin{equation} \Delta\lambda \ge \delta\lambda\left(1-\frac{0.886}{N_{\rm rep}}\right), \label{eq:dlam_pupil} \end{equation}
At $t = 500~\mu$m the requirement falls from $0.0768$ to $0.0670$~nm at $40^\circ$. At larger TIR angle, such as $75^\circ$, the pupil period ($3.73$~mm) exceeds the pupil diameter, so only one replica reaches the eye and the limit vanishes. Thinner substrates admit more replicas ($N_{\rm rep} = 35.8$ at $t = 100~\mu$m for TIR $= 40^\circ$) and so retain a sharp resonance for the modes, reinforcing the thickness dependence of Eqs.~(\ref{eq:theta}) and (\ref{eq:dlambda}). Treating the replicas as equal in amplitude makes Eq.~(\ref{eq:dlam_pupil}) a conservative requirement (Appendix \ref{app:pupil:weights}).

Second, a combiner extracts light on every interaction, so the etalon formed by the two surfaces has finite finesse rather than infinite $Q$ and each guided mode resonance is broadened. With $\eta$ the intensity fraction removed per round trip the amplitude factor is $r=\sqrt{1-\eta}$, the lineshape is Airy, and
\begin{equation} \mathcal{F} = \frac{\pi\sqrt{r}}{1-r} \simeq \frac{2\pi}{\eta} \label{eq:finesse}
\end{equation}
\cite{yariv, ismail}. Resonances separated by the $\delta\lambda$ of Eq.~(\ref{eq:dlambda}) acquire width $\delta\lambda/\mathcal{F}$, so the source need only bridge the remaining gap,
\begin{equation} \Delta\lambda \ge \delta\lambda\left(1-\frac{1}{\mathcal{F}}\right) \simeq \delta\lambda\left(1-\frac{\eta}{2\pi}\right). \label{eq:dlam_loss}
\end{equation}
The approximation works since $\eta$ is very small ($<0.1$). The relaxation is only $\eta/2\pi$, at a representative $\eta = 5\%$ the finesse is $122.5$ and the requirement reduces only by $<1\%$. Out-coupling loss therefore cannot relax the limit of Eq.~(\ref{eq:dlambda}) in practice.

Third, we have considered TE modes only. TM modes form a second, interleaved set whose
positions differ through the polarization dependence of the TIR phase, so illumination
that is unpolarized or that uses both polarizations roughly doubles the density of
supported directions and halves $\delta\theta$. For the cases where laser sources are used and more likely to be constrained by this limit, TE polarization is the most typical polarization.

\section{Conclusion}
\label{sec:conclusion}

Following guided mode theory, we derived the theoretical angular resolution limit of an AR display waveguide, and the two washout mechanisms: source linewidth broadening and thickness non-uniformity. This resolution constraint depends on the thickness, TIR angle, source linewidth, and thickness uniformity. For AR glasses with LED-type displays such as micro-LED displays, the source linewidth is well above $10$~nm, more than two orders of magnitude beyond the $0.1$~nm required at $t = 500~\mu$m, so that the optical resolution is not limited by the guided modes. Independently, local thickness variation across the pupil replication pitch can also remove the limit, but comes with a trade-off with MTF. Surprisingly, pupil truncation removes it only when the pupil replication period approaches the eye pupil diameter, so that a single replica reaches the eye; grating out-coupling loss does not remove it at any practical efficiency. The regime in which the limit genuinely applies is for a thin, uniform waveguide illuminated by a narrow-linewidth laser, and especially for relatively small TIR angles. We show that a waveguide thicker than 1~mm is required to support 1~arcmin optical resolution at TIR $< 42^\circ$. Waveguide design in this case should treat thickness as a resolution parameter alongside the considerations of eyebox uniformity and pupil replication density.
\\

\appendix
\renewcommand{\theequation}{\thesection.\arabic{equation}} \makeatletter \@addtoreset{equation}{section} \makeatother

\renewcommand{\thefigure}{\thesection.\arabic{figure}} \renewcommand{\thetable}{\thesection.\arabic{table}} \makeatletter \@addtoreset{figure}{section} \@addtoreset{table}{section} \makeatother

\textbf{Appendices}
\section{TE guided mode conditions for a symmetric slab}
\label{app:modes}

This appendix derives the eigenvalue conditions quoted as Eqs.~(\ref{eqn15})--(\ref{eqn17}) of
the main text, and the mode spacing $\Delta u = \pi/2$ on which Eq.~(\ref{eq-f}) depends.
The treatment is standard~\cite{yariv} and is reproduced here only to fix notation and to
make the origin of the mode spacing explicit.

\subsection{Wave equation}

Start from Maxwell's equations for a source-free, non-magnetic medium of refractive index
$n$,
\begin{empheq}[left = \empheqlbrace]{align}
\nabla \times \vec{H}&=i \omega \varepsilon_{0} n^{2} \vec{E} \label{a:mx1}\\
\nabla \times \vec{E}&=-i \omega \mu_0 \vec{H} \label{a:mx2}
\end{empheq}
For a mode propagating along $x$ with propagation constant $\beta$,
\begin{empheq}[left = \empheqlbrace]{align}
\vec{H}&=\vec{H}(y, z) \exp [i(\omega t-\beta x)] \\
\vec{E}&=\vec{E}(y, z) \exp [i(\omega t-\beta x)]
\end{empheq}
Eliminating $\vec{H}$ between Eqs.~(\ref{a:mx1}) and (\ref{a:mx2}) gives
\begin{align}
&\nabla \times(\nabla \times \vec{E})=\omega^{2} \varepsilon_{0} \mu_0 n^{2} \vec{E} \\
\Rightarrow\;&\left(\frac{\partial^{2}}{\partial y^{2}}+\frac{\partial^{2}}{\partial z^{2}}\right)
\vec{E}(y ,z)+\left(k_{0}^{2} n^{2}-\beta^{2}\right) \vec{E}(y,z)=0
\label{a:wave}
\end{align}
The slab of Fig.~\ref{fig:1} is invariant along $y$, so $\partial^{2}/\partial y^{2}=0$ and
the field depends on $z$ only. For TE modes the electric field has a $y$ component only,
\begin{equation}
\vec{E}=E_{y}(x, z, t)\,\hat{y}=E(z) \exp [i(\omega t-\beta x)]\,\hat{y}
\end{equation}

\subsection{Guided solutions and boundary conditions}

Solving Eq.~(\ref{a:wave}) subject to decay outside the slab gives
\begin{equation}
\left\{
\begin{aligned}
&E(z)=A \sin (h z)+B \cos (h z), &|z|<t/2\quad\quad\,& \\
&\begin{aligned}
&E(z)=C \exp (-qz), \\
&E(z)=D \exp (+qz), \\
\end{aligned}
&\left.
\begin{aligned}
& z>t / 2 \\
& z<-t / 2 \\
\end{aligned}
\quad\right\}& \text{decaying outside the slab,} \\
\end{aligned}
\right.
\end{equation}
with the transverse wavenumber $h$ inside the core and the decay constant $q$ in the
cladding,
\begin{equation}
\begin{cases}
h^{2}=n_{2}^{2} k_{0}^{2}-\beta^{2} \\
q^{2}=\beta^{2}-n_{1}^{2} k_{0}^{2}
\end{cases}
\label{a:hq}
\end{equation}
Both $E_y$ and the tangential magnetic field
$H_x=-\frac{i}{\omega\mu}\left(\partial E_{y}/\partial z\right)$
must be continuous at $z=\pm\frac{1}{2}t$, which gives four conditions,
\begin{equation}
\begin{cases}
A \sin\left(\frac{1}{2} h t\right) +B \cos \left(\frac{1}{2} h t\right)=C \exp \left(-\frac{1}{2} q t\right) \\
h A \cos \left(\frac{1}{2} h t\right)-h B \sin \left(\frac{1}{2} h t\right)=-q C\exp\left(-\frac{1}{2} q t\right) \\
-A \sin \left(\frac{1}{2} h t\right)+B \cos \left(\frac{1}{2} h t\right)=D \exp \left(-\frac{1}{2} q t\right) \\
h A \cos \left(\frac{1}{2} h t\right)+h B \sin \left(\frac{1}{2} h t\right)=q D \exp \left(-\frac{1}{2} q t\right)
\end{cases}
\end{equation}
Adding and subtracting these in pairs separates the even and odd families,
\begin{equation}
\begin{cases}
2 A \sin \left(\frac{1}{2} h t\right)=(C-D) \exp \left(-\frac{1}{2} q t\right) \\
2 h A \cos \left(\frac{1}{2} h t\right)=-q(C-D) \exp \left(-\frac{1}{2} q t\right) \\
2 B \cos \left(\frac{1}{2} h t\right)=(C+D) \exp \left(-\frac{1}{2} q t\right) \\
2 h B \sin \left(\frac{1}{2} h t\right)=q(C+D) \exp \left(-\frac{1}{2} q t\right)
\end{cases}
\end{equation}

\subsection{Eigenvalue conditions}

For symmetric (even) modes, $A=0$ and $C=D$, and the ratio of the third and fourth lines
gives
\begin{equation}
h \tan \left(\tfrac{1}{2} h t\right)=q .
\label{a:sym}
\end{equation}
For antisymmetric (odd) modes, $B=0$ and $C=-D$, and the first two lines give
\begin{equation}
h \cot \left(\tfrac{1}{2} h t\right)=-q .
\label{a:asym}
\end{equation}
Both families satisfy the single combined relation
\begin{equation}
\tan (h t)=\frac{2 h q}{h^{2}-q^{2}} ,
\end{equation}
obtained from the double-angle identity. Writing $u=\frac{1}{2}ht$ and $v=\frac{1}{2}qt$,
Eq.~(\ref{a:hq}) becomes the circle $u^{2}+v^{2}=\mathbf{V}^{2}$ of Eq.~(\ref{eqn15}), while
Eqs.~(\ref{a:sym}) and (\ref{a:asym}) become $v=u\tan u$ and $v=-u\cot u$, quoted as
Eqs.~(\ref{eqn16}) and (\ref{eqn17}) of the main text.

\subsection{Mode spacing in the thick-slab limit}
\label{app:spacing}

For $t\gg\lambda_0$ the $\mathbf{V}$ circle is large and $v\gg 1$ at every intersection, so
the solutions crowd towards the poles of the two branches. Symmetric solutions require
$u\tan u = v \to \infty$ and therefore approach $u \to m\pi + \pi/2$ from below;
antisymmetric solutions require $-u\cot u = v \to \infty$ and approach
$u \to (m+1)\pi$ from below. Each family is thus spaced by $\pi$ in $u$, and because the two
families interleave, the combined set of guided modes is spaced by
\begin{equation}
\Delta u = \pi/2 ,
\label{a:spacing}
\end{equation}
equivalently $u_m \simeq m\pi/2$ with $m$ integer. This is the result used in
Eq.~(\ref{eq-f}) of the main text, and it is reproduced independently by the
geometric-optics round-trip phase condition of Sec.~\ref{sec:geometric}: with
$2ht-2\phi = 2\pi m$ and $\phi$ slowly varying, $u=\frac{1}{2}ht$ advances by $\pi/2$
between consecutive modes.

\begin{figure}[H]
  \centering
    \includegraphics{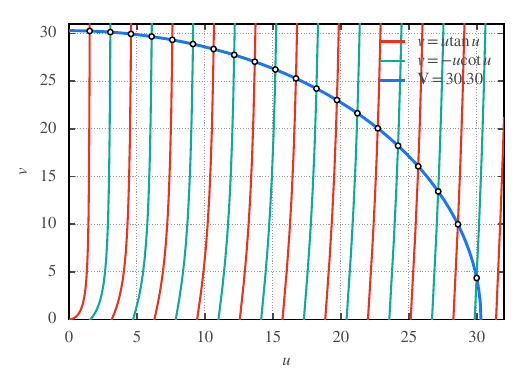}
    \caption{Graphical solution of Eqs.~(\ref{eqn15})--(\ref{eqn17}) for a thin slab. The blue curve is $\mathbf{V}$, and the red branches are for the symmetric modes $v=u\tan u$ while the green branches are for anti-symmetric modes; total 20 modes. $n_1=1$, $n_2=2.7$, $t=2~\mu$m, $\lambda_0=520$~nm, giving $\mathbf{V}=30.30$.}
    \label{fig:app}
\end{figure}

Because $h=\sqrt{n_2^2k_0^2-\beta^2}=n_2k_0\cos\alpha$, each guided mode corresponds to a
distinct TIR angle $\alpha$, and Eq.~(\ref{a:spacing}) therefore fixes the angular spacing
of the modes. Figure~\ref{fig:app} illustrates the graphical solution for a thin slab, where
the individual modes are well separated and can be counted directly.

\section{Mode example with finite light source linewidth}

\begin{figure}[!ht]
  \centering
    \includegraphics{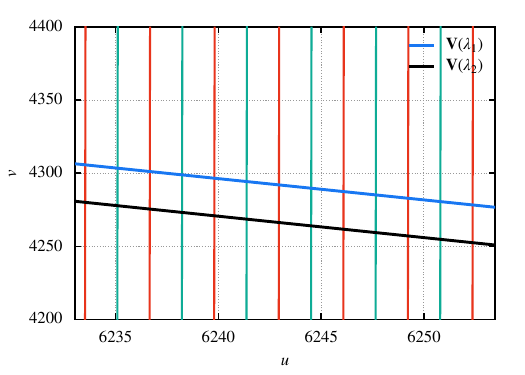}
    \caption{Symmetric (red) and antisymmetric (green) TE modes with $n_1 = 1$, $n_2 = 2.7$, $t = 500~\mu$m, $\lambda_1 = 520$~nm, $\lambda_2 = 521$~nm. Plotted from $u = \frac{1}{2}h_2t$ to $u = \frac{1}{2}h_1t$, where $h_1 = n_2k_1\cos(\alpha_1)$, $h_2 = n_2k_2\cos(\alpha_2)$.}
    \label{fig:mode_linewidth}
\end{figure}

As an example to illustrate conditions described by Eq.~(\ref{eq:dlambda}), using $n_1=1, n_2 = 2.7, \theta=0, \Lambda=300~nm, t = 500~ \mu m$ and the
source spans from $\lambda_1 = 520~ nm$ to $\lambda_2 = 521~ nm$, there are $13$ modes in
total, including symmetric and antisymmetric modes, as shown in
Fig.~\ref{fig:mode_linewidth}. This is a direct check on
Eq.~(\ref{eq:dlambda}): the number of modes within a source bandwidth
$\Delta\lambda$ is $N \approx \Delta\lambda/\delta\lambda = 1~\mathrm{nm}/0.077~\mathrm{nm}
= 13$.

\section{Pupil truncation and the width of the guided-mode resonances}
\label{app:pupil}

The mode picture underlying Eqs.~\eqref{eq:theta} and \eqref{eq:dlambda} assumes an
infinite number of interfering round trips. A real combiner truncates the sum, because the
eye pupil admits only the few out-coupled replicas counted by Eq.~\eqref{eq:Nrep}.
Truncation gives each resonance a finite width, which is what relaxes the linewidth
requirement to Eq.~\eqref{eq:dlam_pupil}. This appendix derives the $0.886/N_{\mathrm{rep}}$
factor, states the criterion it represents, and bounds the error of the two idealisations
it rests on: that $N_{\mathrm{rep}}$ is large, and that the replicas are equally weighted.

\subsection{The finite-replica interference function}
\label{app:pupil:sum}

Write $\delta$ for the phase accumulated in one round trip. Replicas that differ by one
round trip differ in phase by $\delta$, so $N$ replicas of equal amplitude sum as a
geometric series,
\begin{equation}
A(\delta)=\sum_{k=0}^{N-1}e^{\mathrm{i}k\delta}
=\frac{1-e^{\mathrm{i}N\delta}}{1-e^{\mathrm{i}\delta}},
\qquad
I(\delta)=|A|^{2}=\frac{\sin^{2}(N\delta/2)}{\sin^{2}(\delta/2)} .
\label{eq:app:Nslit}
\end{equation}
Equation~\eqref{eq:app:Nslit} is the $N$-beam interference function familiar from the
diffraction grating~\cite{bornwolf}. It peaks at $I=N^{2}$ for $\delta=0 \bmod 2\pi$, and
its period in $\delta$ is $2\pi$, one free spectral range. Close to a peak $\sin(\delta/2)\to\delta/2$, so with $x\equiv N\delta/2$ the envelope of
Eq.~\eqref{eq:app:Nslit} is
\begin{equation}
I \to N^{2}\,\mathrm{sinc}^{2}x , \qquad \mathrm{sinc}\,x=\frac{\sin x}{x} .
\end{equation}
The half-maximum condition is therefore transcendental and has no closed form,
\begin{equation}
\frac{\sin x}{x}=\frac{1}{\sqrt{2}}
\qquad\Longrightarrow\qquad
x_{0}=1.391557\ \mathrm{rad},
\label{eq:app:halfmax}
\end{equation}
giving a full width $2x_{0}$ in $x$, hence $4x_{0}/N$ in $\delta$. As a fraction of the
free spectral range,
\begin{equation}
\boxed{\ \frac{\Delta\delta_{\mathrm{FWHM}}}{2\pi}
=\frac{2x_{0}}{\pi}\,\frac{1}{N}
=\frac{0.886}{N}\ }
\label{eq:app:0886}
\end{equation}
which is the factor used in Eq.~\eqref{eq:dlam_pupil} with $N=N_{\mathrm{rep}}$. The constant
$2x_{0}/\pi$ is the same one that gives the full width at half maximum of the
$\mathrm{sinc}^{2}$ point-spread function of a uniformly illuminated slit,
$0.886\lambda/D$.

\subsection{Validity at small \texorpdfstring{$N_{\mathrm{rep}}$}{Nrep}}
\label{app:pupil:smallN}

Equation~\eqref{eq:app:0886} is a large-$N$ limit. Table~\ref{tab:app:smallN} compares it with the exact half-maximum width of Eq.~\eqref{eq:app:Nslit} with finite integer $N$. The large-$N$ form \emph{understates} the width, by $0.9\%$ at $N=7$ and by $2.7\%$ at $N=4$; it is accurate to better than $0.5\%$ for $N \gtrsim 10$. Since the width enters Eq.~\eqref{eq:dlam_pupil} subtractively, understating it makes the requirement more conservative. Because $N_{\mathrm{rep}}$ from Eq.~\eqref{eq:Nrep} is in general not an integer, the following section treats the non-integer case with non-uniform weights.

\begin{table}[htbp]
\centering
\caption{Exact fractional resonance width of Eq.~\eqref{eq:app:Nslit} against the
large-$N$ form of Eq.~\eqref{eq:app:0886}.}
\label{tab:app:smallN}
\begin{tabular}{rccr}
\hline
$N$ & exact & $0.886/N$ & error \\
\hline
2 & 0.50000 & 0.44295 & $-11.41\%$ \\
3 & 0.31055 & 0.29530 & $-4.91\%$ \\
4 & 0.22770 & 0.22147 & $-2.73\%$ \\
5 & 0.18032 & 0.17718 & $-1.74\%$ \\
7 & 0.12769 & 0.12656 & $-0.88\%$ \\
10 & 0.08897 & 0.08859 & $-0.43\%$ \\
15 & 0.05917 & 0.05906 & $-0.19\%$ \\
20 & 0.04434 & 0.04429 & $-0.11\%$ \\
\hline
\end{tabular}
\end{table}

\subsection{Non-uniform replica weights}
\label{app:pupil:weights}

Equation~\eqref{eq:dlam_pupil} treats the replicas as equal in amplitude. Since each replica is a copy of the in-coupled pupil, and the eye pupil truncates and transmits part of a pupil, full capture requires $|x_{k}|\le(D_{\mathrm{eye}}-D_{\mathrm{ip}})/2$, which for $D_{\mathrm{ip}}=D_{\mathrm{eye}}$ collapses to a single point, so only the exactly centred replica is captured whole in this example case. The relevant amplitude weight is the transmitted \emph{area}, because the far-field amplitude in a given direction is the integral of the pupil field.

For the reference case of the main text, $\lambda_0 = 520$~nm, $n_{2} = 2.7$,
$\Lambda=300$~nm, $\alpha=40^{\circ}$, $t=500$~$\mu$m and
$D_{\mathrm{eye}}=D_{\mathrm{ip}}=3$~mm, the pupil period is $p=0.84$~mm and
Eq.~\eqref{eq:Nrep} gives $N_{\mathrm{rep}}=7.15$. Seven replicas overlap the pupil, with
normalised weights
\begin{equation}
w_{k}=\{0.104,\,0.425,\,0.771,\,1.000,\,0.771,\,0.425,\,0.104\} .
\label{eq:app:weights}
\end{equation}
Evaluating $I(\delta)=|\sum_{k}w_{k}e^{\mathrm{i}k\delta}|^{2}$ gives a width of 0.18973, against 0.12769 for uniform weighting. The clipped resonance is \emph{broader}, because the replicas that contribute are spread over $D_{\mathrm{eye}}+D_{\mathrm{ip}}$ with a taper rather than being uniformly weighted over that span. Applied to Eq.~\eqref{eq:dlam_pupil} with $\delta\lambda=0.0768$~nm, uniform weighting requires $\Delta\lambda\ge0.0670$~nm whereas the clipped weights require only $\Delta\lambda\ge0.0622$~nm.
\emph{ Eq.~\eqref{eq:dlam_pupil} is therefore a conservative bound: it demands more source
linewidth than is actually needed for the washout mechanism.}

\subsection{Single-replica limit}
\label{app:pupil:single}

When $p>(D_{\mathrm{eye}}+D_{\mathrm{ip}})/2$ only one replica overlaps the pupil. There is
then no interference, no mode comb, and no linewidth requirement at all. For
$D_{\mathrm{eye}}=D_{\mathrm{ip}}=3$~mm this threshold is $p>3.0$~mm, reached at
$t=500~\mu$m for $\alpha>72^{\circ}$; at $\alpha=75^{\circ}$,
$p=3.732$~mm and the comb is absent entirely. Equation~\eqref{eq:Nrep} returns
$N_{\mathrm{rep}}=1.61$ there, because $(D_{\mathrm{eye}}+D_{\mathrm{ip}})/p$ is a
span-to-spacing ratio that overcounts once $p$ approaches the span; it should be read as a
replica count only while $p$ is small compared with $D_{\mathrm{eye}}+D_{\mathrm{ip}}$.
Conversely, thinner substrates admit more replicas -- $N_{\mathrm{rep}}=35.8$ at
$t=100~\mu$m -- sharpening the comb and tightening the requirement, which reinforces the
thickness dependence of Eqs.~\eqref{eq:theta} and \eqref{eq:dlambda}.


\bmsection{Acknowledgment}
This work was supported by Reality Labs at Meta. The author would like to thank the colleagues at the Reality Labs research and development teams for helpful discussions.

\bmsection{Disclosures}
The author is an employee of Meta.

\bmsection{Data availability}
No data were generated or analyzed in the presented research. All numerical results reported here are computed directly from the expressions given in the text.

\end{document}